\documentclass{optica-article}

\journal{opticajournal} 

\articletype{Research Article}

\usepackage{lineno}

\date{}
 \usepackage[utf8]{inputenc}
 \usepackage{physics}
 \usepackage{graphicx}
 \usepackage{siunitx}
 \usepackage{hyperref}

 \usepackage{caption}
 \usepackage{subcaption}

\begin{document}

\title{Visualizing strongly focused 3D light fields in an atomic vapor}

\author{Sphinx Svensson,\authormark{1} Clare~R~Higgins,\authormark{2} Danielle Pizzey,\authormark{2} Ifan~G~Hughes,\authormark{2} Sonja Franke-Arnold\authormark{1,*}}

\address{\authormark{1}School of Physics and Astronomy, University of Glasgow, Kelvin Building, Glasgow G12 8QQ, United Kingdom\\
\authormark{2}Department of Physics, University of Durham, Rochester Building, South Road, Durham DH1 3LE, United Kingdom}

\email{\authormark{*}Sonja.Franke-Arnold@glasgow.ac.uk}

\begin{abstract*}
Structured light, when strongly focused, generates highly confined vectorial electromagnetic field distributions which may feature a polarization component along the optical axis. Manipulating and detecting such 3D light fields is challenging, as conventional optical elements and detectors do not interact with the axial polarization component. 
Vector light can, however, be mapped onto atomic polarizations, making electric dipole transitions an ideal candidate to sense such 3D light configurations. Working in the hyperfine Paschen-Back regime, where the electric dipole transitions are spectrally resolved, we demonstrate direct evidence of the axial polarization component of strongly focused radial light. We investigate the influence of various input polarization states, including radial, azimuthal, and higher-order optical vortices, on atomic absorption profiles. Our results confirm a clear mapping between the 3D vector light and the atomic transition strength. This work provides new insights into vectorial light-matter interaction, and opens avenues for novel quantum sensing applications.

\end{abstract*}

\section{Introduction}
The exploration of structured light, shaped in its intensity, phase and polarization, has seen an explosion in research activity over the past decades \cite{RubinszteinDunlop2016}, powered by technological advances in light-shaping techniques \cite{MorenoSLM2020, STAFEEV201732, Forbes2023}, and motivated by a wide range of applications as well as fundamental investigations into geometrical and topological optics. 
The ready availability of light with precisely tailored complex amplitudes structures has unlocked a diverse range of applications, such as advanced microscopy \cite{Chen2023}, optical trapping and manipulation \cite{CURTIS2002169}, and quantum information processing \cite{QuantMem,QuantCommun}. 

Most studies of complex vector light are concerned with transverse phase and polarization features. The familiar paraxial solutions are, however, incompatible with the exact Maxwell equations: all finite-sized light beams contain polarization components along the optical axis \cite{Lax1975}.  These become significant only for non-paraxial beams, where the beam's divergence angle is large \cite{novotny2012principles,zhan2013vectorial,adams2019optics}.  

Light subject to strong focusing is shaped into highly confined complex vectorial electromagnetic field distributions, the shape of which depends on the input polarization state \cite{Huang2011, Bauer2015, Maucher2018, Otte2018, sugic2018singular, Yang2022}. Radially polarized light, in particular, will yield a significant axial polarization component, and can be focused beyond the conventional diffraction limit \cite{youngworth2000focusing, dorn2003sharper} with obvious benefits for superresolution imaging \cite{Chen2019, Kozawa2021}. Further applications include the generation of  optical needles and chains \cite{Wang2008, He2021}; and investigations of nano-plasmonics \cite{Syubaev2019, Hasegawa2024}, and chiral materials \cite{Forbes2023}.
These 3D focal fields are readily modeled, using vectorial diffraction theory, most notably the Richards-Wolf formulation \cite{RW1959}. Recent theoretical work has further explored various aspects of these fields \cite{Cui2018, Andreev2022, MartnezHerrero2023}, including the identification of longitudinal polarization as a prerequisite for transverse spin angular momentum \cite{Aiello2009, Neugebauer2015, Man2020}.

Despite the importance of axial polarization, its direct measurement has proven challenging, as optical components such as polarization optics, photodetectors or CCD cameras are only sensitive to transverse optical fields. Previous attempts to characterize the elusive axial polarization component relied on indirect methods, such as inferring its presence from the fluorescence patterns of single molecules with fixed absorption dipole moments \cite{Novotny2001, Lieb2004}, reconstructing the full 3D field from Mie scattering off a microscopic particle \cite{Bauer2013, Ullah2018}, interaction with molecular monolayers \cite{Otte2019}, or the suppressing of forward scattering when gold nanoparticles were illuminated with radially polarized light \cite{Krasavin2018}. It has also been suggested that the presence of axially polarized light can be detected by observing the spinning and orbiting motions of particles in tightly focused optical tweezers \cite{Zhang2018}. 
Furthermore, generalized Poincar\'e spheres have been introduced to describe e.g.~the directional emission of molecules in arbitrary orientation \cite{Alonso2024}. 

Atoms, however, offer a unique sensitivity to the full 3D electromagnetic field, as electric dipole transitions depend on the alignment between the atomic dipole and the electric field vector. With the optical axis as the quantization axis, superpositions of $\sigma_\pm$ transitions are driven by  polarization transverse to the optical axis, while $\pi$ transitions are driven by axial polarization. This atomic sensitivity has previously been exploited to map transverse optical polarization textures onto the atomic spin textures in paraxial structured vector light \cite{Babiker2018, Wang2020, Wang2024}.
Here, we extend this idea to sense 3D polarization structures via atomic spectroscopy.

Using strongly focused ($\textrm{NA}=0.4$ to $0.7$) complex vector light, we excite $\sigma_\pm$ as well as $\pi$ transitions of the D2 line of $^{87}$Rb in a single-beam geometry. Working in the hyperfine Paschen-Back regime ($1.6$ T magnetic field) allows us to resolve these transitions spectrally.
We present direct evidence of the axial polarization component and verify the spatial polarization distribution of various strongly focused vector beams. We anticipate that this near-resonant atomic detection scheme offers orders of magnitude greater efficiency compared to current 3D field measurement techniques and may provide access to both phase and amplitude information.

Our work provides new insights into magneto-optics: Traditional studies of the interactions of atoms in external magnetic fields with paraxial light fields are typically conducted in two well-known geometrical regimes, the so-called Faraday~\cite{faraday1846experimental} and Voigt ~\cite{voigt1899theorie} geometries, where the incident light wavevector is parallel or perpendicular, respectively, to the magnetic field vector.  Strong focusing enables us to realize combinations of Voigt and Faraday effects simultaneously. This opens exciting possibilities for quantum sensing applications, including magnetometers and magneto-optical filters \cite{uhland2023build, Ilja2014,  Castellucci2021, Ramakrishna2024}.

\section{Polarization structures under strong focusing \label{sec optics}}

Understanding the intricate 3D polarization structures that arise under strong focusing requires a robust theoretical framework. While paraxial light is predominantly polarized within the plane perpendicular to the optical axis, the act of focusing tilts the local wave vectors, and consequently the local polarization. 
Vectorial diffraction theory was developed by Richards and Wolf in 1959 \cite{RW1959} to evaluate the focal fields of homogeneously polarized input light. Today, it provides the necessary tools to model 3D electromagnetic field configurations originating from input light with spatially varying polarization. 

\begin{figure}[!t]
    \centering
    \includegraphics[width = .5\linewidth]
    {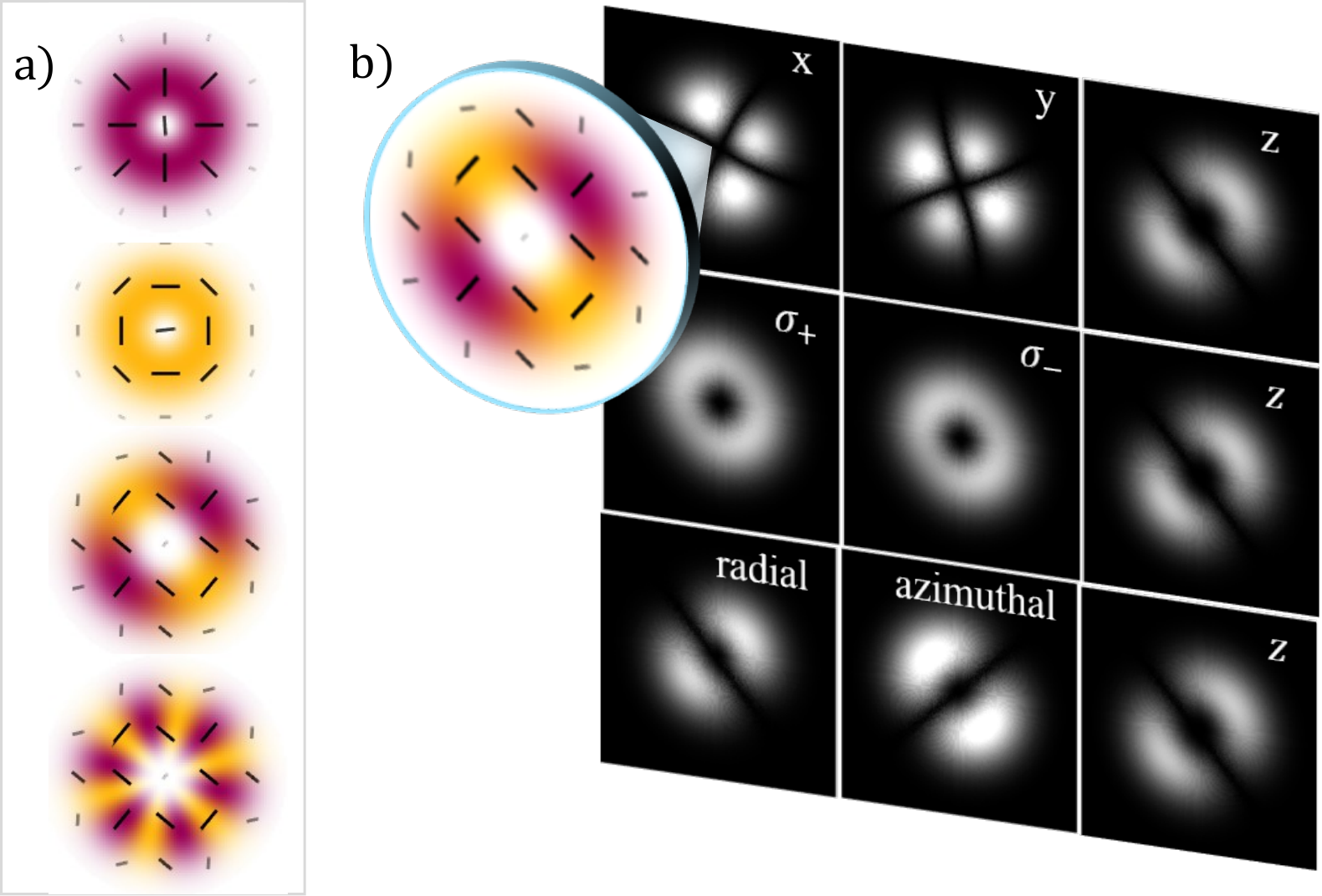}
    \caption[]{Polarization profiles and their simulated 3D structures at the focus. Panel a) shows examples of the investigated polarization structures, color coded in purple for radial and yellow for azimuthal polarization components. From top to bottom: radial $\hat{{\mathbf e}}_\rho$, azimuthal $\hat{{\mathbf e}}_{\phi}$ (Eq.~\ref{eq:etheta}), lemon vortices $\hat{{\mathbf e}}_2$ (Eq.~\ref{eq:e2}) and star vortices $\hat{{\mathbf e}}_6$ (Eq.~\ref{eq:e6}). Panel b) shows the simulated 3D light configurations for $\hat{{\mathbf e}}_2$ at the focus of a lens with NA 0.4. These are displayed in different polarization bases: Cartesian components (top row); circular and axial components corresponding to atomic $\sigma_{\pm}$ and $\pi$ transitions (middle row); and radial, azimuthal and axial polarization (bottom row), highlighting the conversion of some of the radial input polarization into axial polarization.}
    \label{fig: RW}
    \end{figure}

Light beams that are polarized radially and azimuthally with respect to the lens center -- shown as the first two rows of Fig.~\ref{fig: RW}~a) -- display distinct behavior under strong focusing: radial polarization is partially converted into axial polarization (in proportion to the numerical aperture), whereas azimuthal polarization remains azimuthal. Fig.~\ref{fig: RW}~b) illustrates this for a specific input polarization profile.
  
While the Cartesian coordinate system provides a standard description of the electric field, an alternative decomposition into right ($\hat{\sigma}_{-}$) and left ($\hat{\sigma}_{+}$) circular, and axial ($\hat{z}=\hat{\pi}$) polarization components offers a direct connection to atomic transitions.  These circular and axial components are precisely what drive transitions in atoms.  
We note that, while unusual, radial, azimuthal, and axial polarizations provide an alternative valid basis system, which is particularly well suited for visualizing the changes in polarization that occur under strong focusing. In this paper we encode the polarization profiles of the input light in purple for radial and yellow for azimuthal components. 

Building on our discussion of strong focusing effects, we now investigate the specific electromagnetic field configurations created by focusing vector light with a variety of input polarization profiles (Fig.~\ref{fig: RW}\,a.).  These profiles are generated by transmitting linearly polarized input light through a combination of vortex retarders $M_m$ with topological charge $m=1$ or $m=2$, and half-wave plates (HWPs).  The transformations of the paraxial input light are described by the following Jones matrices:

\begin{eqnarray} \label{eq:Jones}
M_{m}=\begin{pmatrix}
 \cos(m \phi)& \sin(m \phi) \\
\sin(m \phi) & -\cos(m \phi) 
\end{pmatrix}, \quad \mathrm{HWP}_\theta= \begin{pmatrix}
 \cos(2 \theta)& \sin(2 \theta) \\
\sin(2 \theta) & -\cos(2 \theta) 
\end{pmatrix},
\end{eqnarray}

where $\phi$ is the azimuthal angle within the beam, and $\theta$ defines the orientation of the HWP's fast axis with respect to the horizontal. 
The vortex retarders introduce an on-axis polarization singularity by manipulating the differential phase between orthogonal polarization components of the input beam.
In the far field however, after spatial filtering, the intensity structure approximates that of the corresponding eigenmodes of propagation, namely the Laguerre-Gauss mode $\mathrm{LG}_0^m$. 

To achieve rotationally symmetric polarization structures with varying ``radiality'', we transmit horizontally polarized light through a HWP followed by an $m=1$ vortex retarder.  This process yields a polarization structure:
\begin{eqnarray} \hat{{\mathbf e}}_\theta= M_{1}\cdot \mathrm{HWP}_\theta \cdot \begin{pmatrix} \label{eq:etheta}
1 \\ 0 \end{pmatrix} 
=  \begin{pmatrix}\cos(2\theta-\phi)\\ -\sin(2\theta-\phi) \end{pmatrix}.
\end{eqnarray}
Rotating the HWP from $\theta=0$ to $\theta=\pi/4$ changes the polarization profile gradually from radial, $\hat{{\mathbf e}}_\rho=\hat{{\mathbf e}}_0$, to azimuthal $\hat{{\mathbf e}}_\phi=\hat{{\mathbf e}}_{\pi/4}$.

Polarization profiles with a two-fold symmetry (and a lemon singularity) or a six-fold symmetries (and a star singularity)\footnote{For a recent review on topological defects see \cite{Wang2021}.} can be generated by an $m=2$ vortex retarder followed by a HWP at $\theta=0$  or $\theta=\pi/8$, respectively: 
\begin{eqnarray} \label{eq:e2}\hat{{\mathbf e}}_2 & = & \mathrm{HWP}_0 \cdot M_{2}\cdot \begin{pmatrix}
1 \\ 0 \end{pmatrix} 
= \begin{pmatrix}\cos(2\phi)\\ -\sin(2\phi) \end{pmatrix},  \\ \label{eq:e6}
\hat{{\mathbf e}}_6 & = & \mathrm{HWP}_{\pi/8} \cdot M_{2}\cdot \begin{pmatrix}
1 \\ 0 \end{pmatrix} 
= \frac{1}{\sqrt{2}} \begin{pmatrix}\sin(2\phi) + \cos(2\phi)\\ \cos(2\phi) - \sin(2\phi) \end{pmatrix}.
\end{eqnarray}
(The HWP preceding the retarder may remain in place, set to $\theta=0$ so as not to modify the horizontal input polarization.)

\section{Experimental setup}
\begin{figure*}
    \centering
    \includegraphics[width = 0.8\linewidth]{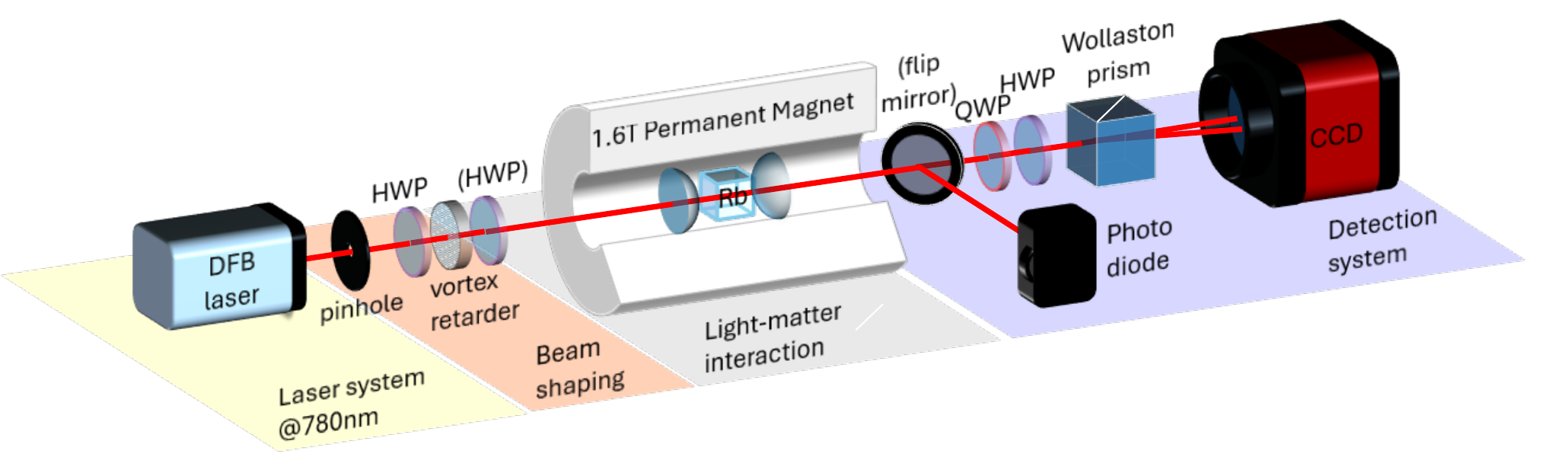}
    \caption[]{Simplified diagram of the experimental setup. Light at $780\,$nm is generated by a DFB laser and scanned over 100 GHz. Various vector beam profiles can be generated by a sequence of vortex retarders and waveplates. This vector beam is strongly focused, and interacts with a warm rubidium vapor within an external magnetic field of 1.6 T. Its absorption spectrum can be measured with a photodiode; alternatively its spatially resolved intensity and polarization profile can be captured on a CCD camera after Stokes polarimetry.   
}
    \label{fig:setup}
\end{figure*}

The experimental setup, outlined in Fig. \ref{fig:setup}, is designed to investigate the interaction of strongly focused vector vortex light with $^{87}$Rb vapor in a strong magnetic field. A distributed-feedback laser (DFB) generates $780\,$nm light, tunable over 100 GHz, encompassing all transitions in the hyperfine Paschen-Back regime. The laser profile is spatially filtered through a 50 $\upmu$m pin hole to homogenize its intensity profile and its initial polarization defined by a polarizing beam splitter cube (not shown).
The polarization profile is then set by a sequence of a half-wave plate (HWP), a vector vortex retarder ($m=1$ or $m=2$), and an optional additional HWP (see Section~\ref{sec optics}). The shaped light is focused into a 1 mm$^{3}$ glass cell containing isotopically enriched, heated $^{87}$Rb vapor, typically operated at temperatures between 75 -- 85$^{\circ}$C. Outside the focal region the ring of highest intensity for the radial and azimuthal beams is 4.0~mm (for the two-fold and six-fold beams 5.7~mm), while at the focus it is 0.4~$\upmu$m (0.57~$\upmu$m), and peak intensity of 0.005~$\upmu$W/m$^2$ (0.0025~$\upmu$W/m$^2$). The cell and focusing lenses (NA = 0.4 or 0.7) are housed within a Teflon/aluminium unit, positioned at the center of a NdFeB permanent magnet~\cite{trenec2011permanent}, producing a uniform 1.6 T magnetic field along the optical axis. Spectroscopic data can be acquired using a photodiode and spatially resolved spectroscopy on a CCD camera. Additionally, the polarization profile of the light can be captured by spatially resolved full Stokes tomography, by placing polarization optics and a Wollaston prism before the CCD camera. 

The 1.6 T magnetic field ensures operation well within the hyperfine Paschen-Back (HPB) regime, where the Zeeman splitting dominates the ground state hyperfine interaction (Fig.~\ref{fig:1.6T}). A more detailed analysis of the level structure is given in Section 1 in the Supplemental Material. This separation of Zeeman-induced transitions exceeds the Doppler width, enabling observation of distinct $\sigma_\pm$ and $\pi$ transitions (Fig.~\ref{fig:1.6T}).

\begin{figure}
    \centering
    \includegraphics[width=.9\linewidth]{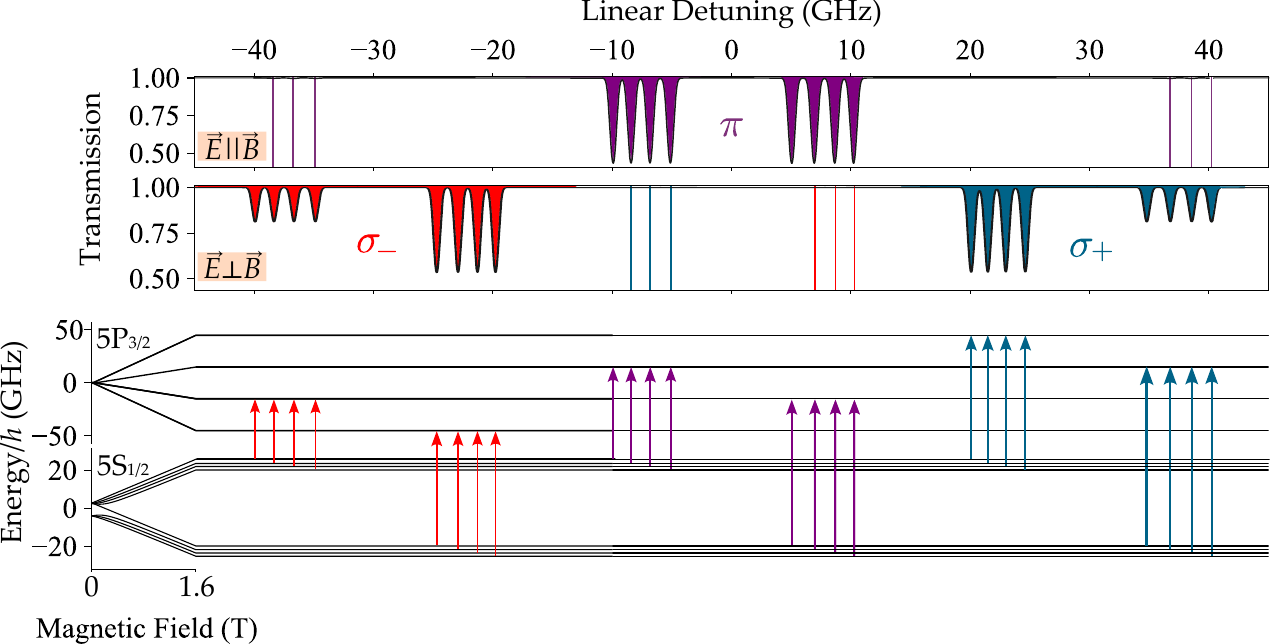}
    \caption{Energy levels involved in the $^{87}$Rb D$_{2}$ transitions in the presence of a 1.6~T external magnetic field (lower panel) and corresponding simulated transmissions (top panel). When $\Vec{E} \parallel \Vec{B}$, corresponding to the Voigt configuration, linearly polarized light excites $\pi$ transitions. When $\Vec{E} \perp \Vec{B}$, corresponding to the Faraday configuration, linearly polarized light  excites superpositions of $\sigma_{-}$ (red) and  $\sigma_{+}$ (blue) transitions. Possible (much weaker) transitions due to ground state admixtures are indicated by lines in the transmission spectrum. The spectra were simulated using a computational modeling tool -- $ElecSus$ \cite{Zentile2015b} -- for an atomic vapor cell of length 1~mm at a temperature of 75$^{\circ}$C.}
    \label{fig:1.6T}
    \end{figure}

Precise alignment is critical in this experiment: the center of the (approximately) Gaussian intensity profile of the input beam should coincide with the singularity of the polarization vortex imposed by the vortex retarders, as well as with the optical axis of the high NA imaging system (pre-aligned within the magnet's core). Furthermore, the optical axis should be aligned with the magnetic field direction which we take as the quantization axis of the atoms. 
Any misalignment between the optical axis of the input light and that of the imaging system introduces an effective tilt of the optical propagation direction with respect to the quantization axis, thereby generating unwanted artificial contributions to the $\pi$ transition. 

\section{Experimental Results and Discussions}

In our first experiment we visualize the presence of an axial polarization component in strongly focused light (NA\,=\,0.4) through HPB absorption spectroscopy. The effective NA is slightly smaller as we do not overfill the lens, and its variation across the beam profile is taken into account by the numerical analysis. 
We monitor the strength of the $\pi$ transitions with a photodiode in the far field of the Rb sample, and verify the predicted dependence on the input polarization profile. 

\begin{figure}
    \centering
    \includegraphics[width = .8\linewidth]
    {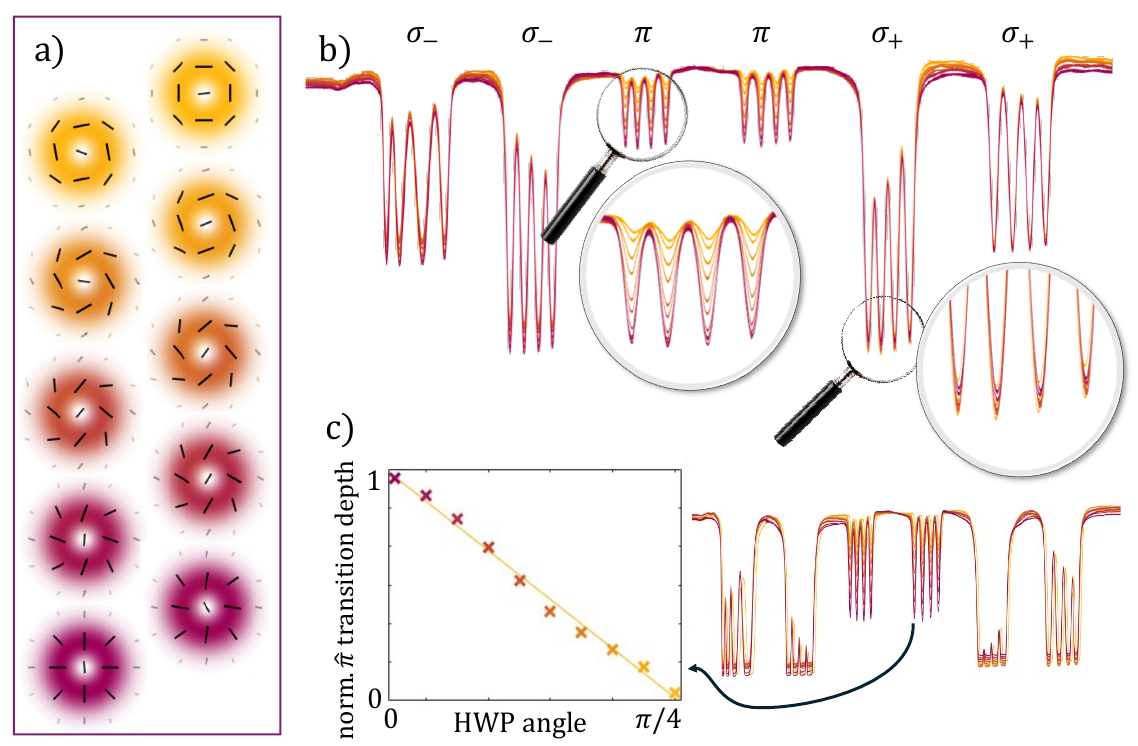}
    \caption[]{Evidence of a varying longitudinal polarization component as function of the ``radiality'' of the light beam. Absorption spectra of the D2 transition for Rb for input beams shown in panel a) with various polarization profiles from azimuthal (yellow) to radial (dark red). The spectra in b) for Rb heated to approximately \SI{80}{\celsius} feature $\sigma_\pm$ as well as $\pi$ transitions. The $\pi$ transition absorption is smallest for azimuthally polarized light beams and becomes largest for radial polarization.  Panel c) shows the depth of the leftmost $\pi$ transition peak on the right block as a function of the beams ``radiality'' from the equivalent dataset taken at $\approx$~125$^\circ$C. }\label{fig:AbsorptionSpectra}
\end{figure}

According to Richards \& Wolf diffraction theory \cite{RW1959, youngworth2000focusing}, the electric field components at the focus are a function of both the NA and the input polarization. An azimuthal input field remains purely transverse throughout the entire focal region, and hence should not be able to excite the $\pi$ transition. A radial input field, in contrast, generates a longitudinal component, whose strength depends on the apodization function resulting from the intensity profile of the input light and the NA of the lens. The longitudinal polarization component is hence directly proportional to the radial component of the input beam. To test this, we vary the polarization profile of the input light gradually from azimuthal to radial as detailed in Section \ref{sec optics}. By rotating the HWP before the $M_1$ vortex retarder in 10$^\circ$ steps we generate beams with polarizations corresponding to the simulated profiles shown in Fig.~\ref{fig:AbsorptionSpectra}\,a). For each of these beam profiles we record 10 complete absorption traces by scanning the DFB laser over 100~GHz, which we then normalize to the off-resonance intensity in the vicinity of the $\pi$ transition, and average to produce the absorption spectra shown in Fig.~\ref{fig:AbsorptionSpectra}b) and c).
The frequency scan, controlled by the DFB laser's temperature, is not linearized or referenced to an atomic standard in zero magnetic field; therefore, absolute frequency calibration is not possible \cite{pizzey2022laser}. A detailed discussion of the atomic spectroscopy is provided in the Supplemental Material. 

The absorption spectrum, shown in Fig.~\ref{fig:AbsorptionSpectra}, contains  both $\sigma_\pm$ and $\pi$ transition, usually indicative of measurements in the Faraday and Voigt configuration respectively. 
The $\pi$ transitions are minimal for the azimuthal input beam, but increase in strength as the input polarization becomes more radial. As the longitudinal beam component grows, the transverse components decreases, manifest in slightly diminished $\sigma_\pm$ transitions. 
Fig.~\ref{fig:AbsorptionSpectra}b) was taking at a vapor temperature of approximately 80$^\circ$C. 

To quantify the dependence of transition strength on the polarization profile, we increased the optical density by heating the cell to $\approx$~125$^\circ$C, and analyze the transition strength of the leftmost $\pi$ transition as a function of the beam's radial component as set by the HWP angle during beam preparation. Fig.~\ref{fig:AbsorptionSpectra}\,c) shows the transition strength, normalized against that for purely radial input light.\footnote{\,This transition is chosen due to the absence of ground state admixture. See Supplemental Material for more information.}  
As predicted, the observed transition strength depends linearly on the radial component within the input light, which in turn is proportional to its longitudinal polarization at the focus. The small residual absorption dips for the azimuthal beam are attributed to previously discussed alignment imperfections and slight input beam inhomogeneity. Even a small angle between the external magnetic field and the average propagation direction will induce $\pi$ absorption.

We note that even homogeneous linearly polarized light, when strongly focused, will feature longitudinal polarization components, which can be observed as absorption of $\pi$ transitions. We present evidence of this in Section 2 of the Supplemental Material.

While we have thus far considered only the overall presence of the longitudinal light component, our experimental setup enables us to explore the spatial distribution of the different polarization components, revealing the 3D electric field's shape at the focus. As indicated in Fig.~\ref{fig: RW}, we expect the longitudinal polarization component, and hence the absorption at the $\pi$ transition, to mirror the radially polarization component of the input beams.

To investigate these spatial variations, we test polarization profiles with different rotational symmetry (shown in Fig.~\ref{fig: RW}~a): a radially and an azimuthally polarized beam as well as polarization profiles with two-fold and six-fold symmetry as defined in Eqs.~\ref{eq:etheta}, \ref{eq:e2} and \ref{eq:e6}.  
We record the spatial absorption profiles across the full 100~GHz frequency scan on a CCD camera.  
These measurements were performed with a NA~=~0.4 lens to facilitate alignment of the beams through the optical system and the rubidium vapor cell. The transmitted beams were recorded at 30 frames per second, with each full scan comprising 304 images, and averaged over 2--4 scans per dataset.

\begin{figure*}
    \centering
    \includegraphics[width = .8\linewidth,trim=0 0cm 0 0cm]{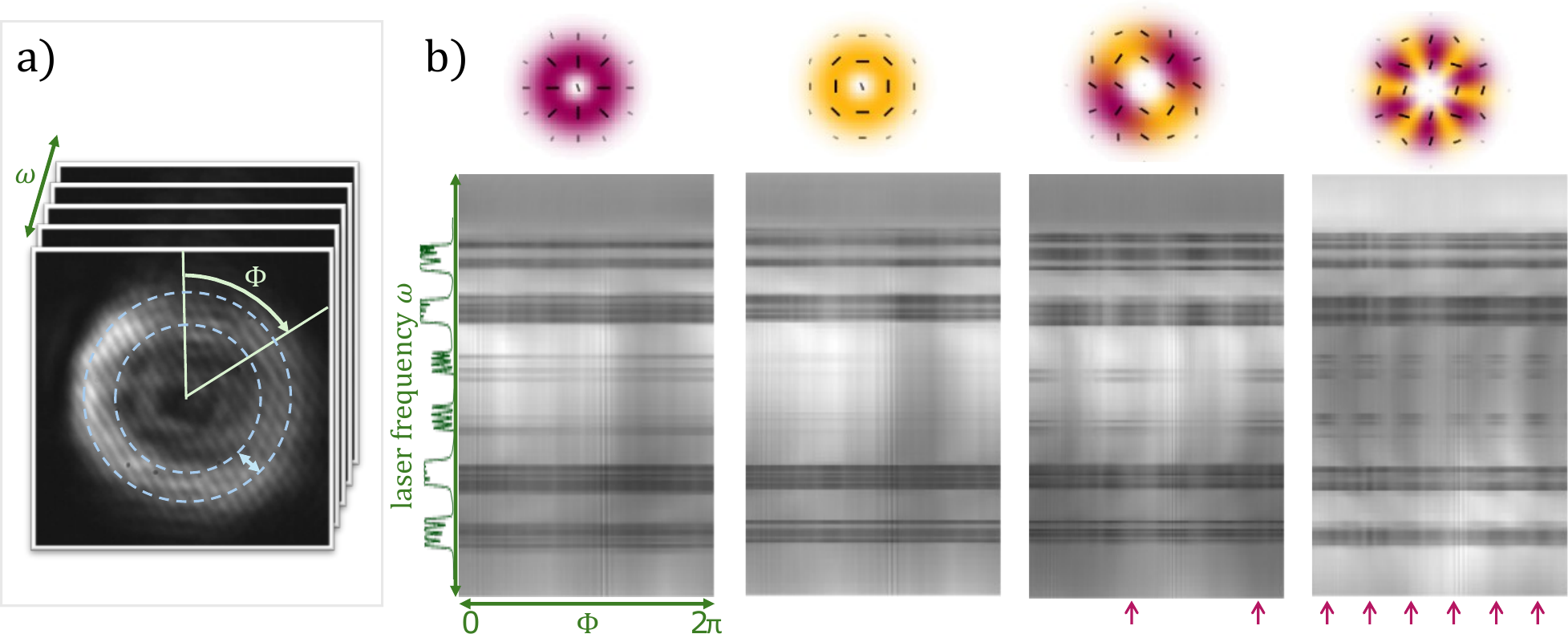}
    \caption[]{Spatially resolved absorption spectra for strongly focused (NA~=~0.4) vector light beams with various radial symmetries.  Panel a) shows the conversion of spatially resolved spectroscopy videos into the visual representation in b), as function of laser frequency $\omega$ and azimuthal angle $\Phi$ (averaged over the indicated ring of width $\Delta r$).  The resulting absorption maps in b) depict, from left to right, the response to radial polarization $\hat{{\mathbf e}}_\rho$, azimuthal $\hat{{\mathbf e}}_{\phi}$, lemon vortices $\hat{{\mathbf e}}_2$ and star vortices $\hat{{\mathbf e}}_6$. The $\sigma_\pm$ transitions are present in all maps at all azimuthal angles, whereas the $\pi$ transition matches the geometry of the polarization profile: It is present for radial light, much fainter for azimuthal light, and is interrupted twice and six times for $\hat{{\mathbf e}}_2$ and $\hat{{\mathbf e}}_6$, respectively, indicated by the red arrows.  
    }
    \label{fig: unwrappy intplots}
    \end{figure*}
We initially investigate the azimuthal symmetry of the observed absorption profiles in Fig.~\ref{fig: unwrappy intplots}. Fig.~\ref{fig: unwrappy intplots}~a) illustrates the conversion of recorded spectroscopy videos, shown as a stack of absorption images, into the data presented in  Fig.~\ref{fig: unwrappy intplots}~b). For each input polarization we display the intensity as a function of laser frequency $\omega$ and unwrapped azimuthal angle $\Phi$, averaged over the brightest region of the beam indicated by the dashed ring (using an intensity threshold of 43\% of the peak intensity).
A polynomial fit to the off-resonance azimuthal intensity distribution is used to normalize the beam profile for the entire frequency scan. Each row along the vertical axis corresponds to a frame from the spectroscopy video. As expected, the $\sigma_{\pm}$ transitions are rotationally uniform for all input beams, see Supplemental Material Fig.~S4. The structure of the $\pi$ transition, however, varies significantly: uniform absorption is observed for the radial beam, while it is nearly absent for the azimuthal beam. For input beams with varying degrees of radiality, the strength of the $\pi$ absorption is modulated with the same symmetry as the input polarization structure, consistent with vector diffraction theory. Light containing a polarization vortex $\hat{{\mathbf e}}_6$ with six alternating regions of azimuthal and radial symmetry, for example, shows six angular regions where the $\pi$ transitions are present, indicating six angular regions where the focused light contains longitudinal polarization components.

\begin{figure*}
    \centering
    \includegraphics[width = .8\linewidth]{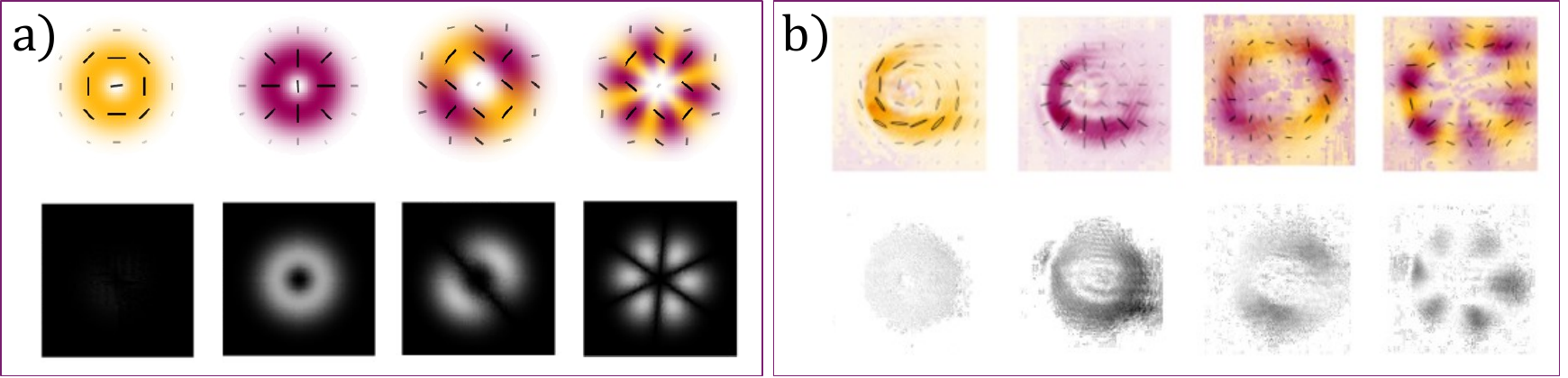}
    \caption[]{Visualizing the axial component of various strongly focused (NA~=~0.4) vector light beams via atomic optical densities at the $\pi$ transition, illustrated for azimuthal and radial polarizations as well as higher order vortices with two-fold and six-fold symmetry. Panel a) shows the theoretical polarization profiles and axial intensity distributions at the focus for these four beams. Panel b) shows the corresponding experimental polarization patterns as well as the optical density of the atomic vapor at one of the $\pi$ transitions.   }
    \label{fig:intensityvids}
\end{figure*} 
The connection between the polarization structure and the atomic transition strength becomes even more evident when considering the full spatial information of the frames corresponding to the $\pi$ transition. In Fig.~\ref{fig:intensityvids}, we show direct evidence of the spatial structure of the longitudinal polarization component in the absorption images taken at the same $\pi$ transition as considered in Fig.~\ref{fig:AbsorptionSpectra}\,b). For each beam, the simulated intensity structure of its longitudinal component (Fig.~\ref{fig:intensityvids}\,a) aligns with areas of radial polarization. Fig.~\ref{fig:intensityvids}\,b) presents comparisons between the measured input polarization profiles (top row, obtained via Stokes tomography) and the corresponding optical density\footnote{\,Optical density is calculated as the ratio of the transmitted to input light: OD = $\frac{I_{\mathrm{trans}}}{I_{\mathrm{in}}}$.} at a laser frequency resonant with the leftmost $\pi$ transition (bottom row). Given that all input beams are linearly polarized, they contain equal amounts of right and left circular polarization across their profiles. Consequently, the absorption images at the $\sigma_{\pm}$ transitions are also rotationally symmetric, as shown in the Supplemental Material (Fig.~S4).

\section{Conclusion}

We have demonstrated a novel approach to probing the intricate 3D polarization structure of light using atomic spectroscopy in the hyperfine Paschen-Back regime. Specifically, we have shown, for the first time, simultaneous spectroscopy of the $\sigma_{\pm}$ and $\pi$ transitions in a single-beam setup, directly observing the axial polarization component inherent in strongly focused vector beams. Our observation of $\pi$ transition excitation by radially polarized light, and its absence with azimuthal polarization, provides compelling evidence of this axial component.  Furthermore, we have shown how the spatial absorption profile of structured light is modified depending on the input polarization. By driving the normally ``forbidden'' $\pi$ transition in a Faraday configuration, we have directly measured this axial component -- a feat previously unrealized. This work validates the predictions of Richards \& Wolf theory and opens new avenues for manipulating atomic states with structured light. The ability to selectively control atomic transitions through input polarization offers unprecedented control over information encoded in atomic systems, with potential applications ranging from enhanced optical devices to novel quantum sensing techniques.  Critically, this work lays the foundation for exploring new phenomena related to partial absorption and the interaction of structured light with atomic media, including polarization-dependent focusing effects and the study of longitudinal polarization's birefringent and dichroic properties.
~\\

\begin{backmatter}
\bmsection{Funding}
SF-A and SS acknowledge support through the QuantERA II Programme, with funding received via the EU H2020 research and innovation programme under Grant No. 101017733 and associated support from EPSRC under Grant No. EP/Z000513/1 (V-MAG). IGH acknowledges the funding received from EPSRC (Grant No. EP/R002061/1).

\bmsection{Acknowledgments}
The authors gratefully acknowledge the assistance of Hannah Frost and the Mechanical Engineering Services within the Physics Department at Durham University for their expertise in designing and machining a custom mount for the vapor cell and high numerical aperture lenses. The authors would like to thank Jacques Vigu\'e for kindly providing the 1.6~T permanent magnet. We are grateful for illuminating discussions with Niclas Westerberg, Richard Maduro and Kuntal Samanta within the School of Physics and Astronomy at the University of Glasgow.

\bmsection{Disclosures}
The authors declare no conflict of interest.

\medskip

\bmsection{Data availability} Data underlying the results presented in this paper are available in XXX

\end{backmatter}

\clearpage %
\appendix 
\section*{Supplementary Material} 


\title{Visualizing strongly focused 3D light fields in an atomic vapor: supplemental document}
\author{} 




\begin{abstract}
This document is supplemental material for the paper ``Visualizing strongly focused 3D light fields in an atomic vapor''. Throughout this document, we reference equations and figures from the paper using \emph{Eq.~(X)} and \emph{Fig.~X} respectively.
\end{abstract}




\section{Hyperfine Paschen-Back regime}

In the presence of a magnetic field, the energy levels of alkali-metal atoms undergo splitting due to the interaction between the magnetic field and the atom's magnetic dipole moment.  At low magnetic fields, the total angular momentum $\mathbf{F} = \mathbf{I} + \mathbf{J}$ is a good quantum number, where $\mathbf{I}$ is the nuclear spin and $\mathbf{J}$ is the total electron angular momentum.  However, as the magnetic field strength increases, the interaction of $\mathbf{I}$ and $\mathbf{J}$ with the external field becomes stronger than their mutual interaction.  This leads to a decoupling of $\mathbf{I}$ and $\mathbf{J}$, and the projections $m_I$ and $m_J$ become good quantum numbers.  This regime, where the Zeeman shift exceeds the ground-state hyperfine splitting, is known as the hyperfine Paschen-Back (HPB) regime.

The magnetic field strength required to reach the HPB regime is characterized by $B_{\rm HPB}~=~A_{\rm hf}/\mu_{\rm B}$, where $A_{\rm hf}$ is the ground-state magnetic dipole constant and $\mu_{\rm B}$ is the Bohr magneton.  For $^{87}$Rb, this characteristic field is approximately 0.23 T~\cite{sargsyan2022saturated}. All results reported in this work are taken at $1.6\,$T, well within the HPB regime.

Spectroscopy of alkali-metal atoms in the HPB regime has been extensively investigated~\cite{windholz1985zeeman, windholz1988zeeman, tremblay1990absorption, umfer1992investigations, PhysRevA.84.063410, sargsyan2017decoupling, mathew2018simultaneous, keaveney2019quantitative, staerkind2023precision, mottola2023electromagnetically, mottola2023optical, Briscoe_2024, staerkind2024high, higgins2024fine}.  The Breit-Rabi diagram~\cite{breit1931measurement, Woodgate1980} illustrates the evolution of energy levels as a function of the applied magnetic field (see Fig.~3 in the main text).  In the HPB regime, the Zeeman splitting between dipole-allowed transitions exceeds the Doppler width, resulting in a spectrum of isolated atomic resonances. 

\begin{figure}
    \centering
    \includegraphics[width=.8\linewidth]{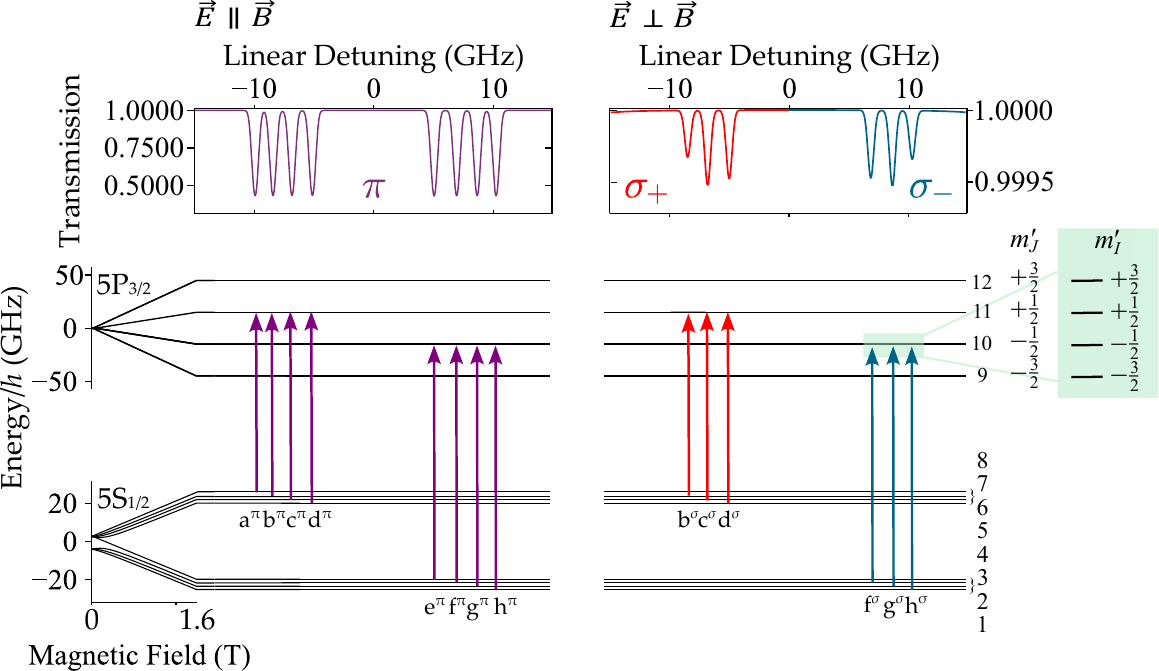}
       \caption{Energy levels involved in the $^{87}$Rb D$_{2}$ transitions in the presence of a 1.6~T external field. The simulated absorption spectrum of linearly polarized light when $\Vec{E} \parallel \Vec{B}$, which excites $\pi$ transitions, is shown at the left, while the simulated absorption spectrum of linearly polarized light when $\Vec{E} \perp \Vec{B}$, which excites $\sigma_{-}$ (red) and  $\sigma_{+}$ (blue) transitions is shown at the right. The spectra were simulated using a computational modeling tool -- $ElecSus$ \cite{Zentile2015b} -- for an atomic vapor cell of length 1~mm at a temperature of 75$^{\circ}$C.}
    \label{fig:big_diagram_theory}
\end{figure}

Furthermore, the simplified energy level structure in the HPB regime facilitates the comparison of experimental spectra with theoretical models~\cite{WhitingEIA, WhitingEIT,  whiting2017single, whiting2018four, higgins2021electromagnetically}. This isolation of transitions is particularly advantageous for our experiment, as it allows us to excite selectively  the $\pi$ transitions, which are otherwise inaccessible in the Faraday geometry. For paraxial light beams, $\pi$ transitions are only accessible in the Voigt (or skewed Voigt) geometry, where the magnetic field has a component perpendicular to the light propagation direction. However, by tightly focusing our incident light in the HPB regime, we generate a significant longitudinal polarization component, enabling the excitation of $\pi$ transitions even in the Faraday geometry, i.e.~for a magnetic field  parallel to the light propagation. This phenomenon is illustrated in Fig.~\ref{fig:big_diagram_theory}.
The simulated spectra depict the distinct $\pi$ transitions (left column) and $\sigma_{\pm}$ transitions (right) column, corresponding to the electric field vector parallel and perpendicular to the magnetic field, respectively. For completeness we give the admixtures of the various ground states in Table \ref{tab:my_label}. 

\begin{table}[h!]
\small
    \centering
    \begin{tabular}{ll|ll}
    \hline
        8: & ${\color{violet} 1.0000 \ket{+\frac{1}{2},+\frac{3}{2}}}$ & 4: & ${\color{violet} 1.0000 \ket{-\frac{1}{2},-\frac{3}{2}}}$  \\
        7: & {\color{violet} 0.9979 $\ket{+\frac{1}{2},+\frac{1}{2}}$} + {\color{teal} 0.0647 $\ket{-\frac{1}{2},+\frac{3}{2}}$} & 3: & {\color{violet} 0.9971 $\ket{-\frac{1}{2},-\frac{1}{2}}$} $-$ {\color{red} 0.0759 $\ket{+\frac{1}{2},-\frac{3}{2}}$}  \\
        6: & {\color{violet}0.9968 $\ket{+\frac{1}{2},-\frac{1}{2}}$} + {\color{teal}0.0805 $\ket{-\frac{1}{2},+\frac{1}{2}}$} & 2: & {\color{violet} 0.9968 $\ket{-\frac{1}{2},+\frac{1}{2}}$} $-$ {\color{red} 0.0805 $\ket{+\frac{1}{2},-\frac{1}{2}}$} \\
        5: & {\color{violet}0.9971 $\ket{+\frac{1}{2},-\frac{3}{2}}$} + {\color{teal} 0.0759 $\ket{-\frac{1}{2},-\frac{1}{2}}$} & 1: & {\color{violet} 0.9979 $\ket{-\frac{1}{2},+\frac{3}{2}}$} $-$ {\color{red} 0.0647 $\ket{+\frac{1}{2},+\frac{1}{2}}$} \\
        \hline
    \end{tabular}
    \caption{Admixtures of the ground states $\ket{m_J,m_I}$, color coded according to the supported transitions.}
    \label{tab:my_label}
\end{table}

\section{Strong focusing effects for linearly polarized light}
\begin{figure}[!ht]
    \centering
    \includegraphics[width=0.6\linewidth]{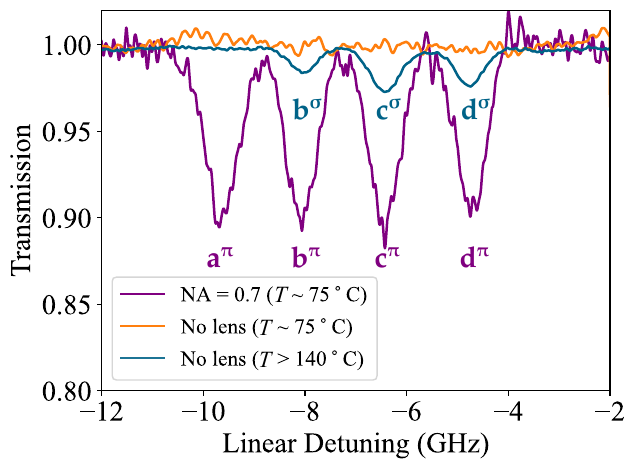}
    \caption{The experimental absorption spectrum of circularly polarized light through an atomic vapor cell of length 1~mm in the presence of a 1.6~T magnetic field in Faraday geometry. With no high-NA, the $\sigma^{+}$ transitions -- labeled b$^{\sigma}$, c$^{\sigma}$, and d$^{\sigma}$, to coincide with the labeling in Supplementary Fig.~\ref{fig:big_diagram_theory} -- become apparent at temperatures much greater than 140$^{\circ}$C, where the medium is optically thick (blue). At a lower temperature of 75$^{\circ}$C, the $\sigma^{+}$ transitions are not visible (orange). With the addition of a NA = 0.7 lens pair, the $\pi$ transitions -- labeled a$^{\pi}$, b$^{\pi}$, c$^{\pi}$ and d$^{\pi}$, to coincide with the labeling in Fig.~\ref{fig:big_diagram_theory} -- are excited at $T$~=~75$^{\circ}$C (purple).}
    \label{fig:big_diagram_exp_data}
\end{figure}

Linearly polarized light contains alternating areas of radial and azimuthal polarization components, and hence is predicted to produce longitudinal polarization, when strongly focused. In preliminary experiments we have verified the impact of strong focusing on the transition strengths of linearly polarized light.
Fig.~\ref{fig:big_diagram_exp_data} compares the spectroscopy signal for a quasi paraxial beam, when no lenses are used, and for a lens pair with NA~=~0.7. 
When high-NA lenses are used at an atomic vapor temperature of 75$^\circ$C, we observe excitation of both $\pi$ (shown as purple curve in Fig.~\ref{fig:big_diagram_exp_data}) and $\sigma_{\pm}$ transitions, confirming the presence of the longitudinal polarization component. For paraxial light at the same vapor temperature these transitions vanish (orange curve). If the interaction strength is increased by raising the vapor temperature above 140$^\circ$C, $\sigma_{\pm}$ transitions can be observed, while the longitudinal component remains insufficient to excite the $\pi$ transitions. Notably, we observe three $\sigma_{\pm}$ transitions, compared to the four $\pi$ transitions, in agreement with the simulated spectrum of Fig.~\ref{fig:big_diagram_theory} . This difference arises because one of the $\pi$ transitions (a$^{\pi}$, as labeled in Fig.~\ref{fig:big_diagram_theory}) lacks ground-state hyperfine admixture, making it a clean transition for monitoring strong focusing effects. The same applies to the transition e$^{\pi}$, and we use it for experiments on structured light reported in the main document in Fig.~4c).

\section{Alignment procedure}

The conversion of polarization vortices of the input light into 3D polarization structures at the focus requires a precise alignment between the polarization vortex and the optical axes, as well as between the optical axes and the imaging system. The former is achieved by accurate positioning of the vortex retarder with respect to the Gaussian beam profile. 
Here we detail our alignment procedure for the latter.

\begin{figure}[!ht]
    \centering
    \includegraphics[width = .8\linewidth]
    {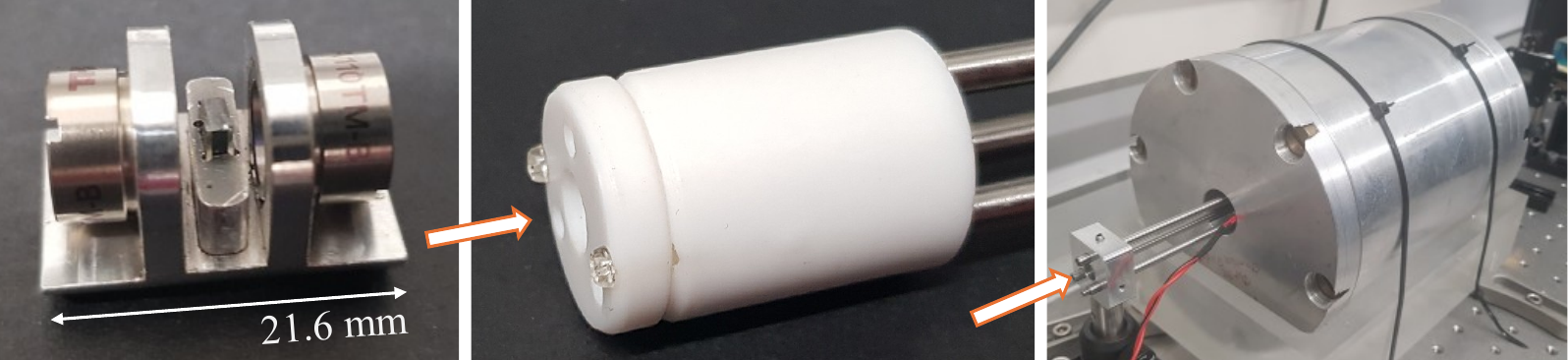}
    \caption[]{From left to right:  aluminium holder with two plano-convex lenses separated by $2 f$ (Thorlabs A110TM-B, $f=6.24\,$mm and NA~=~0.40) and the $1\,$mm vapor cell; 3D printed holder with alignment rods, allowing to position the imaging system at the core of a NdFeB permanent magnet. }
    \label{fig: photos}
    \end{figure}

Alignment is achieved in three steps: 
While the bore in the magnet is still empty, two 3D printed alignment aids, with centered holes of 3~mm diameter, are fitted on either end of the magnet, allowing us to align the beam through the bore and center it on the camera, thereby setting the optical axis. In the second alignment step, the custom-designed imaging system (Fig.~\ref{fig: photos}) is aligned with the optical axis. An aluminum holder containing the pre-aligned lenses and heatable vapor cell at their focus is positioned in a 3D-printed holder and inserted into the center of the magnet. Its position and orientation is adjusted via alignment rods and optimized for minimal beam displacement and distortion, controlled on the camera. In the final stage, the alignment is optimized by minimally altering the input beam position and angle to achieve the best beam shape. At this stage an azimuthal beam can be used to optimize the oscilloscope signal for a minimal $\pi$ transition, or alternatively a radial beam can be used to check if the $\pi$ transition evenly affects the outside of the beam. Once alignment and optimization are complete, the experiment can begin.

\section{Spectroscopy} 

A DFB laser scans its frequency by altering its temperature, which is controlled by an external signal generator set to a triangle wave. However, the temperature of the laser is not linearly related to the output wavelength, and therefore the frequency scan is nonlinear. In order to provide a scale for the frequency axes later on, some fits to other datasets and simulations were performed, though these are of course associated with large errors. The data presented in this paper has not been altered to linearize the scans.

The DFB laser does not produce a Gaussian beam, and is therefore spatially filtered by a 50 $\mu$m pinhole. This increases, however, the inherent intensity fluctuations significantly over the course of a scan, so the atoms will be subjected to additional spatial and temporal intensity fluctuations.

In an early version of the cell holder, the cell walls were fully perpendicular to the propagation axis, however this caused the cell to act as a cavity, broadening and distorting the signal. This effect was mitigated in the current model of the cell holder by placing the cell at a tilt to the propagation axis, which brought on new challenges: the tilted glass panes of the cell refract the beam at different distances from the focus, changing its spatial structure and propagation direction. A distortion of the three-dimensional beam shape in and after the cell is inevitable, in part responsible for the residual $\pi$ transition observed even for azimuthal input beams in Figs.~4, 5 and 6 of the main text.

The oscilloscope traces shown in Fig.~4 in the main document 
were taken in 100 consecutive measurements on the same day. For this, the beam exiting the atoms was focused into a photodiode and the resulting intensities were displayed in a time resolved manner on an oscilloscope. For each of the 10 polarization profiles, set by the HWP angle, 10 scans were taken and averaged. 

Spatially resolved absorption spectra (paper Figs.~5 and 6) were acquired in the form of a video of the beam after the atoms. The frame rate of the CCD camera was 30 Hz, and each frame corresponds to a detuning range of roughly $330\,$MHz. In order to identify the image frames corresponding to the individual $\sigma_\pm$ and $\pi$ transitions, we summed over the total intensity in each image, thereby recreating absorption spectra as obtained via oscilloscope traces. 

\section{Additional data sets}

\begin{figure}[!ht]
    \centering
    \includegraphics[width = .9\linewidth]
    {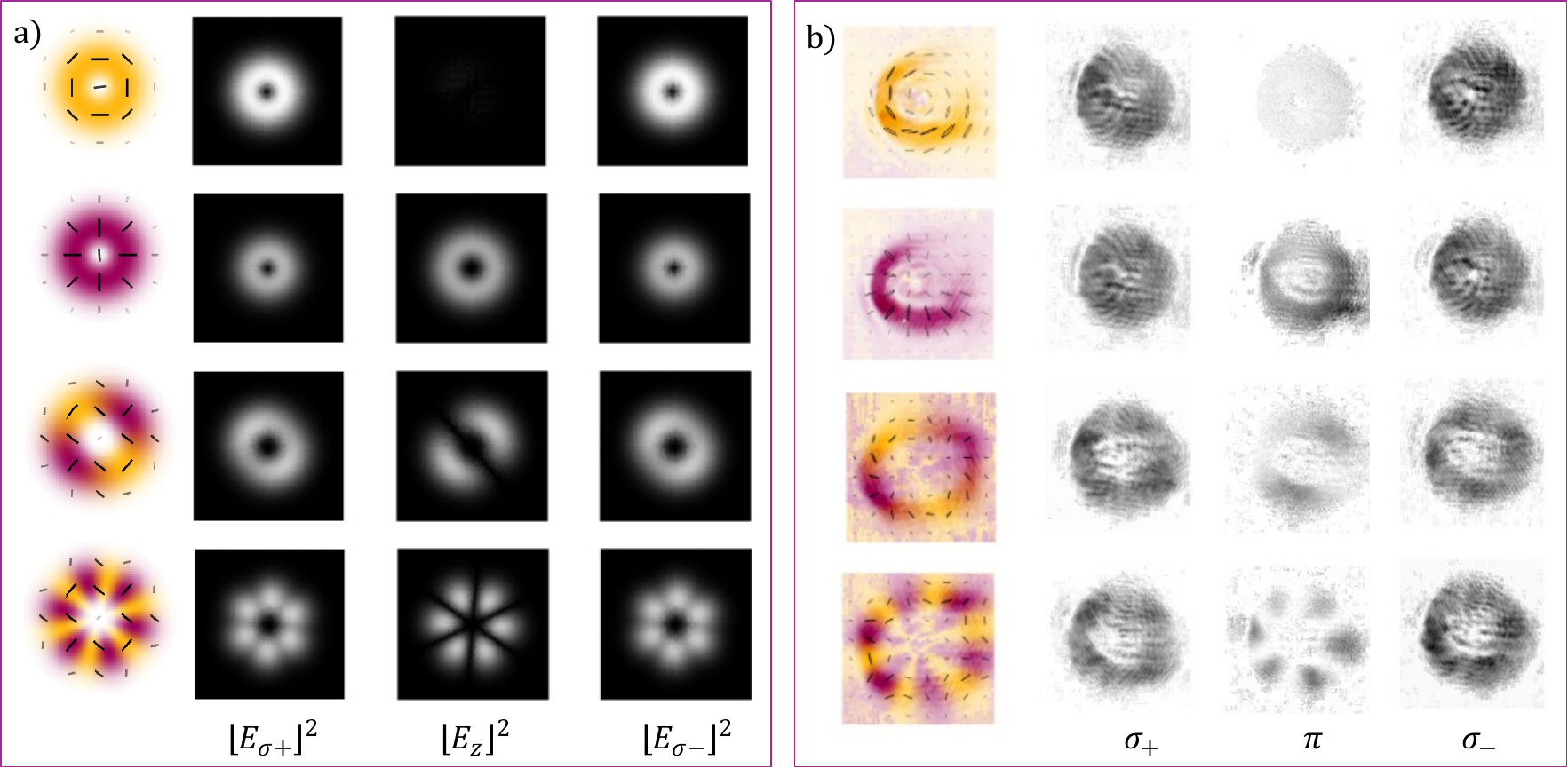}
    \caption[]{Mapping the 3D electric field at the focus of various strongly focused (NA~=~0.4) vector light beams via atomic optical densities at the $\sigma_+$, $\pi$ and $\sigma_-$ transition, illustrated for azimuthal and radial polarizations as well as higher order vortices with two-fold and six-fold symmetry. Panel a) shows the theoretical polarization profiles of the input beams, and the corresponding simulated intensity contributions at the focus for $\hat{\sigma}_+$, $\hat{z}$ and $\hat{\sigma}_-$ polarization components. Panel b) shows the corresponding experimental polarization patterns as well as the optical density of the atomic vapor at one of the $\sigma_+$, $\pi$ and $\sigma_-$ transition. }
    \label{fig: OD_all}
    \end{figure}



\clearpage

\bibliography{bib}

\end{document}